\documentclass[letterpaper,twocolumn,aps,prl,showpacs]{revtex4-1}
\usepackage[latin9]{inputenc}
\usepackage{verbatim,color}
\usepackage{amsmath,txfonts}
\usepackage{amssymb}
\usepackage{graphicx}
\usepackage{hyperref}


\begin{document}

\title{Soft superconducting gap in semiconductor Majorana nanowires}
\author{So Takei}
\affiliation{Condensed Matter Theory Center and Joint Quantum Institute, Department
of Physics, University of Maryland, College Park, Maryland 20742-4111,
USA.}
\author{Benjamin M. Fregoso}
\affiliation{Condensed Matter Theory Center and Joint Quantum Institute, Department
of Physics, University of Maryland, College Park, Maryland 20742-4111,
USA.}
\author{Hoi-Yin Hui}
\affiliation{Condensed Matter Theory Center and Joint Quantum Institute, Department
of Physics, University of Maryland, College Park, Maryland 20742-4111,
USA.}
\author{Alejandro M. Lobos}
\affiliation{Condensed Matter Theory Center and Joint Quantum Institute, Department
of Physics, University of Maryland, College Park, Maryland 20742-4111,
USA.}
\author{S. Das Sarma}
\affiliation{Condensed Matter Theory Center and Joint Quantum Institute, Department
of Physics, University of Maryland, College Park, Maryland 20742-4111,
USA.}
\date{\today}
\begin{abstract}
We theoretically consider the ubiquitous soft gap measured in the tunneling conductance of semiconductor-superconductor hybrid structures, in which recently observed signatures of elusive Majorana bound states have created much excitement.
We systematically study the effects of magnetic and non-magnetic disorder, temperature, dissipative Cooper pair breaking, and interface inhomogeneity, which could lead to a soft gap. We find that interface inhomogeneity with moderate dissipation is the only viable mechanism that is consistent with the experimental observations. Our work indicates that improving the quality of the superconductor-semiconductor interface should result in a harder induced gap.
\end{abstract}
\pacs{73.63.Nm, 74.45.+c, 74.81.-g, 03.65.Vf}
\maketitle

\textit{Introduction.} The pursuit of exotic topological phases of
matter has become an exciting topic of research in physics~\cite{kitaev2001}.
In particular, topological superconductors (SCs) supporting zero-energy
Majorana bound states (MBS)~\cite{Nayak08_RMP_Topological_quantum_computation,DasSarma_PRB'06,tewari_prl'2007,Fu08_Proximity-effect_and_MF_at_the_surface_of_TIs,zhang_prl'08,Sato09_Topological_phases_in_noncentrosymmetric_SC,Sau10_Proposal_for_MF_in_semiconductor_heterojunction,Lutchyn10_MF_and_Topological_transition_in_SM_SC_Heterostructures,Oreg10_Helical_liquids_and_MF_in_QW,Alicea10_MF_in_tunable_semiconductor_devices,Jiang2011,RokhinsonNP}
have received increasing attention both for its intrinsic interest and for its potential uses in topological
quantum computation~\cite{Nayak08_RMP_Topological_quantum_computation,sauetalTQC,aliceaetalnature,beenakkerNJP}.
In recent proposals to realize topological SCs in solid state systems,
an effective $p$-wave SC is induced in a semiconductor by the
combined effects of spin-orbit coupling (SOC), Zeeman splitting
of the energy bands and proximity induced $s$-wave SC~\cite{Sau10_Proposal_for_MF_in_semiconductor_heterojunction,Lutchyn2010,Oreg2010}. 
For a semiconductor nanowire (NW), it was predicted that the presence of MBS at its ends 
could be experimentally detected in the differential tunneling conductance $G(V)$
at the interface with a normal contact, via
the emergence of a zero-bias peak (ZBP) of height $2e^{2}/h$ (at
zero temperature)~\cite{SenguptaPRB,Law,Sau_long, Flensberg,Wimmer}. The NW proposal has 
inspired a number of recent experiments in which 
suggestive ZBPs 
have been observed~\cite{Mourik12_Signatures_of_MF, Das12_Evidence_of_MFs, Deng12_ZBP_in_Majorana_NW,Finck,Chang}.
However, whether these ZBPs are truly due to Majorana zero modes is still uncertain.
In particular, while it has been
argued that disorder could lead to a spurious non-topological ZBP in the experiments ~\cite{Bagrets12_Class_D_spectral_peak_in_Majorana_NW,Liu12_ZBP_in_Majorana_wires_with_and_without_MZBSs,Pikulin12_ZBP_from_weak_antilocalization_in_Majorana_NW},  it has been recently suggested that (contrary to the common expectation) disorder does not necessarily destroys the topological phase in proximity-induced SC NWs, and therefore the observed ZBPs could in principle have a topological origin \cite{Adagideli13_Topological_order_in_dirty_wires}.

%

An ubiquitous feature of all Majorana experiments involving proximity-induced superconductivity 
has remained ignored in the literature despite a great deal of activity in the field: the measured 
$G(V)$ is extremely ``soft" in both the high-field topological 
phase (where the ZBP exists) and in the zero-field or the 
low-field trivial phase (where there is no ZBP).  In fact, 
the soft gap feature, which is clearly a property of the semiconductor-SC hybrids quite independent of the MBS physics, 
is prominent in the data with the subgap conductance 
being typically only a factor of 2-3 lower than the above-gap conductance, 
implying the existence of rather large amount of subgap states whose origin 
remains unclear. We believe that without a thorough understanding of this ubiquitous
soft gap, our knowledge of the whole subject remains incomplete.

In this Letter, we develop a minimal theoretical model that may generally 
explain the soft gap that is observed ubiquitously in the current Majorana 
experiments~\cite{Mourik12_Signatures_of_MF, Das12_Evidence_of_MFs, Deng12_ZBP_in_Majorana_NW, Finck}.
 We systematically consider the effects due to: (a) non-magnetic and (b) magnetic disorder
in the NW; (c) temperature; (d) dissipative quasiparticle broadening 
arising due to various pair-breaking mechanisms such as poisoning, coupling to other degrees of freedom (e.g. phonons or normal electrons in the leads)
or due to electron-electron interactions; and (e) inhomogeneities
at the SC-NW interface due to imperfections (e.g. roughness and barrier fluctuations) that may arise during
device fabrication. Since the soft gap occurs universally in the experiment at all parameter values, we consider only the 
non-topological zero-magnetic field situation here because this is where the gap should be the largest 
and the hardest. We solve our model numerically by exact diagonalization of the Hamiltonian, and complement the study using the Abrikosov-Gor'kov formalism \cite{Abrikosov60_SC_alloys_with_impurities} for a simplified model of a semiconductor NW with a spatially-fluctuating pairing potential. 
%
%

Our results point to the inhomogeneities at the semiconductor NW-SC interface [i.e. mechanism (e)] as the main physical mechanism producing the soft gap.
Our work indicates that improving the quality of the superconductor-semiconductor interface should result in a harder induced gap and in a simpler physical interpretation of the Majorana experiment.
However,  our conclusions are not restricted to Majorana NWs and might be useful for a correct interpretation of the experimental results in many semiconductor-SC hybrid systems.
%

\textit{Theoretical model.}
We consider a one-dimensional semiconductor NW of length $L_x$ placed along the $x$-axis and subjected to SOC,
Zeeman field along its axis, and proximity-induced $s$-wave pairing due to a proximate bulk SC.
Discretization of the Hamiltonian in the continuum results in a tight-binding model defined on a $N_{x}$-site lattice \cite{Stanescu11_MFs_in_SM_nanowires},
\begin{flalign}
&\hat{H}_{w} =-t\sum_{\langle ij\rangle,s}c_{is}^{\dagger}c_{js}+i\alpha\sum_{i,ss'}
\left[c^\dag_{i s}\sigma^y_{ss'}(c_{i+1 s'}-c_{i-1 s'})\right]  \nonumber\\
&-\sum_{i,ss'}c_{is}^{\dagger}\left[\mu_i-B_{Z}\sigma^{x}-\boldsymbol{b}_{i}\cdot\boldsymbol{\sigma}\right]c_{is'}  
 +\sum_i \left[\Delta_{i}c_{i\uparrow}^{\dagger}c_{i\downarrow}^{\dagger}+{\rm h.c.}\right].\label{eq:lattHam}
\end{flalign}
Here, $c_{is}^{\dagger}$ creates an electron with spin $s=\uparrow,\downarrow$ at site $i$, 
$\alpha=\alpha_R/2a=\sqrt{E_{so}t}$ parametrizes the Rashba SOC strength, where $E_{so}=m_{e}^{*}\alpha_{R}^{2}/2$ 
is the SOC energy scale, $\alpha_{R}$ is the Rashba velocity and $a$ is the lattice constant. 
$B_{Z}$ is the Zeeman energy, and $\boldsymbol{\sigma}=(\sigma_{x},\sigma_{y},\sigma_{z})$
is the vector of Pauli matrices. We use for the NW $L_x=N_xa=2 \mu$m, $m_e^*=0.015 m_e$, $E_{so}=50 \mu $eV, 
and temperature $T=70{\rm mK}$~\cite{Mourik12_Signatures_of_MF}. We assume a one-band model with $N_x =500$, 
$t=676\mu$eV, and $\alpha = 0.07 t$.

Static non-magnetic disorder in the NW is included through a fluctuating chemical potential
$\mu_i=\mu_0+\delta\mu_i$ around the average value $\mu_0$. Static magnetic disorder may be present 
in the sample due to contamination with magnetic atoms or due to the presence of regions in the NW acting
as quantum dots with an odd number of electrons. 
Here, we neglect the quantum dynamics of the impurity spins and model its effect as a randomly
oriented inhomogeneous magnetic field $\boldsymbol{b}_i$~\cite{balatsky_RevModPhys'06}.

The effects of the proximate bulk SC on the NW are modeled  in Eq. (\ref{eq:lattHam}) by an effective {\em locally}-induced {\em hard} gap $\Delta_i$. The locality of the induced pairing interaction is justified because the coherence length of the bulk SC is typically much shorter ($\xi_\text{SC}\approx 3$nm in NbTiN alloys) than the Fermi wavelength of the semiconductor NW ($\lambda_\text{F}\approx 10^2$nm). The assumption of an induced hard  gap is justified if the SC-NW interface is in the tunneling regime. This seems to be a reasonable assumption since the experimentally reported induced gaps are much smaller than the parent bulk SC gaps \cite{Mourik12_Signatures_of_MF, Das12_Evidence_of_MFs, Deng12_ZBP_in_Majorana_NW}, a fact that typically occurs in low-transmittance interfaces \cite{mcmillan68_proximity_effect, Aminov96_SN_interfaces, Stanescu11_MFs_in_SM_nanowires} (As a word of caution, the experimental evidence for this identification is still 
limited and other explanations cannot be completely ruled out).  In the tunneling regime, the quantity  $\gamma_i=\rho_0t^2_{\perp,i}\ll\Delta_{\rm SC}$, 
where $\rho_{0}$ is the local density of states of electrons in the NW at the Fermi energy in the normal phase, $t_{\perp,i}$ is
the local tunneling matrix element at the NW-SC interface at site $i$, and $\Delta_\text{SC}$ the bulk parent gap in the SC. Then, the bulk SC is known to
induce a hard gap in the NW, 
$\Delta_i\approx\gamma_i$~\cite{mcmillan68_proximity_effect, Stanescu11_MFs_in_SM_nanowires}.
A more general treatment of the SC-NW interface that takes into account higher orders in $t_{\perp,i}$ (i.e., highly transparent interfaces) is outside the scope of the present work, and we refer the reader to the well-known bibliography on the subject \cite{Blonder1982,Neurohr1996,Volkov1993}. 

Inhomogeneities at semiconductor-SC interfaces are known to occur generically due to sample fabrication procedures, and their effects have been extensively studied
 (see e.g. Refs.~\onlinecite{Huffelen1993,Neurohr1996}). In our model, we take into account these inhomogeneities through local spatial fluctuations in $t_{\perp,i}$, which effectively give rise to spatial fluctuations in the induced $s$-wave SC pairing $\Delta_i$ in Eq. (\ref{eq:lattHam}).
We assume $t_{\perp,i}=t_{\perp}^0e^{-\kappa\delta d_i}$,
where $\delta d_i$ denotes the fluctuation in the width of the
NW-SC barrier and $\kappa$ is a phenomenological
constant with units of inverse length that parametrizes the energy barrier of the
NW-SC interface. Such a functional form is expected due to fluctuations in the overlap of evanescent wavefunctions. Then, the induced SC pairing is  $\Delta_i=\Delta_{0}e^{-2 \delta\beta_i}$, where the
dimensionless parameter $\delta\beta_i=\kappa\delta d_i$
characterizes the roughness of the interface, and $\Delta_0$ is the induced SC
pairing in the absence of the interface inhomogeneity (we take the value $\Delta_0 = 250 \mu $eV from Ref. ~\onlinecite{Mourik12_Signatures_of_MF}).  
Note that our model for interface fluctuations is generic and only incorporates the inevitable presence of potential fluctuations 
at the interface separating the SC metal and the NW. 

The different disorder mechanisms are taken into account by introducing
Gaussian-distributed random variables $\delta\mu_{i}$, $\boldsymbol{b}_{i}=\left(b_{i}^{x},b_{i}^{y},b_{i}^{z}\right)$,
and $\delta\beta_{i}$ with zero means and variances given by $\left\langle \delta\mu_{i}\delta\mu_{j}\right\rangle =W_{\mu}^{2}\delta_{ij}$,
$\left\langle b_{i}^{p}b_{j}^{q}\right\rangle =W_{b}^{2}\delta_{ij}\delta_{pq}$,
and $\left\langle \delta\beta_{i}\delta\beta_{j}\right\rangle =W_{\beta}^{2}\delta_{ij}$,
respectively. To model the interface inhomogeneity, we coarse-grain the interface in patches of length $5a$ and assume 
that $\delta\beta_i$ is uniform within each patch, but varies randomly from patch to patch with a standard
deviation of $W_\beta$. Note that assuming a Gaussian distribution
in $\delta\beta_{i}$ results in a different probability distribution
function for $\Delta_i$
\begin{equation}
P\left(\Delta_i\right)  =\frac{1}{2\Delta_i\sqrt{2\pi}W_{\beta}}\textrm{exp}\left[-\frac{1}{8W_{\beta}^{2}}\textrm{ln}^{2}\left(\frac{\Delta_i}
{\Delta_0}\right)\right].\label{eq:p_Delta}
\end{equation}

The relevant experimental quantity is the
tunneling differential conductance $G\left(V\right)$ at an end of the
NW, which is related to the local density of states~\cite{Mourik12_Signatures_of_MF, Das12_Evidence_of_MFs, Deng12_ZBP_in_Majorana_NW,Chang}.
We calculate $G\left(V\right)$ using the tunneling formalism by coupling the NW to a contact lead~\cite{meir92,Sau_long,Flensberg}.
The Hamiltonian of the combined system is $\hat{H}=\hat{H}_{w}+\hat{H}_{L}+\hat{H}_{t}$,
where 
$\hat{H}_{L}=\sum_{ks}\varepsilon_{k}d_{ks}^{\dagger}d_{ks}$ is the
Hamiltonian describing the lead and $\hat{H}_{t}=t_{L}\sum_{ks}d_{ks}^{\dagger}c_{1s}+\text{h.c.}$
is the tunneling Hamiltonian coupling site $i=1$ of the NW to the lead via a tunneling matrix element $t_{L}$.
The tunneling conductance at site $i=1$ reads
\begin{equation}
G\left(V,T\right) =-2\pi e^{2}t_{L}^{2}\rho_{L}\int_{-\infty}^{\infty}d\omega\rho_{1}^{w}\left(\omega\right)f'\left(\omega-eV\right),
\label{eq:tunneling_conductance}
\end{equation}
where $f\left(x\right)$ is the Fermi
distribution function, $\rho_{L}$ is the lead density of states at the Fermi energy, and $V$ is the voltage at which the lead is biased with respect to $\mu_0$. 
Here, $\rho_{1}^{w}\left(\omega\right)$ is
the local density of states in the NW (including both spin projections) at site $i=1$ in the presence
of the lead, which we calculate as $\rho_{i}^{w}\left(\omega\right)=-\frac{1}{\pi}\text{Im }g_{ii}^{w}\left(\omega\right)$.
 Here $g_{ij}^{w}\left(\omega\right)$ is the retarded Green's function of the NW in real-space representation, which in the limit $t_{L}\rightarrow0$ becomes
\begin{equation}
g_{ij}^{w}\left(\omega\right) =\sum_{ns}\frac{u_{is,n}^{\left(0\right)*}u_{js,n}^{\left(0\right)}}{\omega-E_{n}^{\left(0\right)}+i\gamma_{L,n}}+\frac{v_{is,n}^{\left(0\right)*}v_{js,n}^{\left(0\right)}}{\omega+E_{n}^{\left(0\right)}+i\gamma_{L,n}},
\label{eq:greens_function}
\end{equation}
with $E_{n}^{\left(0\right)}$ and $\{u_{is,n}^{\left(0\right)},v_{is,n}^{\left(0\right)}\}$
being, respectively, the eigenvalues and eigenvectors resulting
from the diagonalization of the BdG Hamiltonian corresponding
to Eq.~(\ref{eq:lattHam}). To include the presence of the lead, we
solve the equation of motion for $g_{ij}^{w}\left(z\right)$ in the
presence of $\hat{H}_{t}$ \cite{mahan2000}. The term $\gamma_{L,n}$ is the self-energy, which in the limit $t_{L}\rightarrow0$
becomes  
$\gamma_{L,n}=-i\pi\rho_{L}t_{L}^{2}\sum_{s}\left(|u_{1s,n}^{\left(0\right)}|^{2}+|v_{1s,n}^{\left(0\right)}|^{2}\right)$ .

\begin{figure}[t]
\centering 
\includegraphics[width=0.4\textwidth]{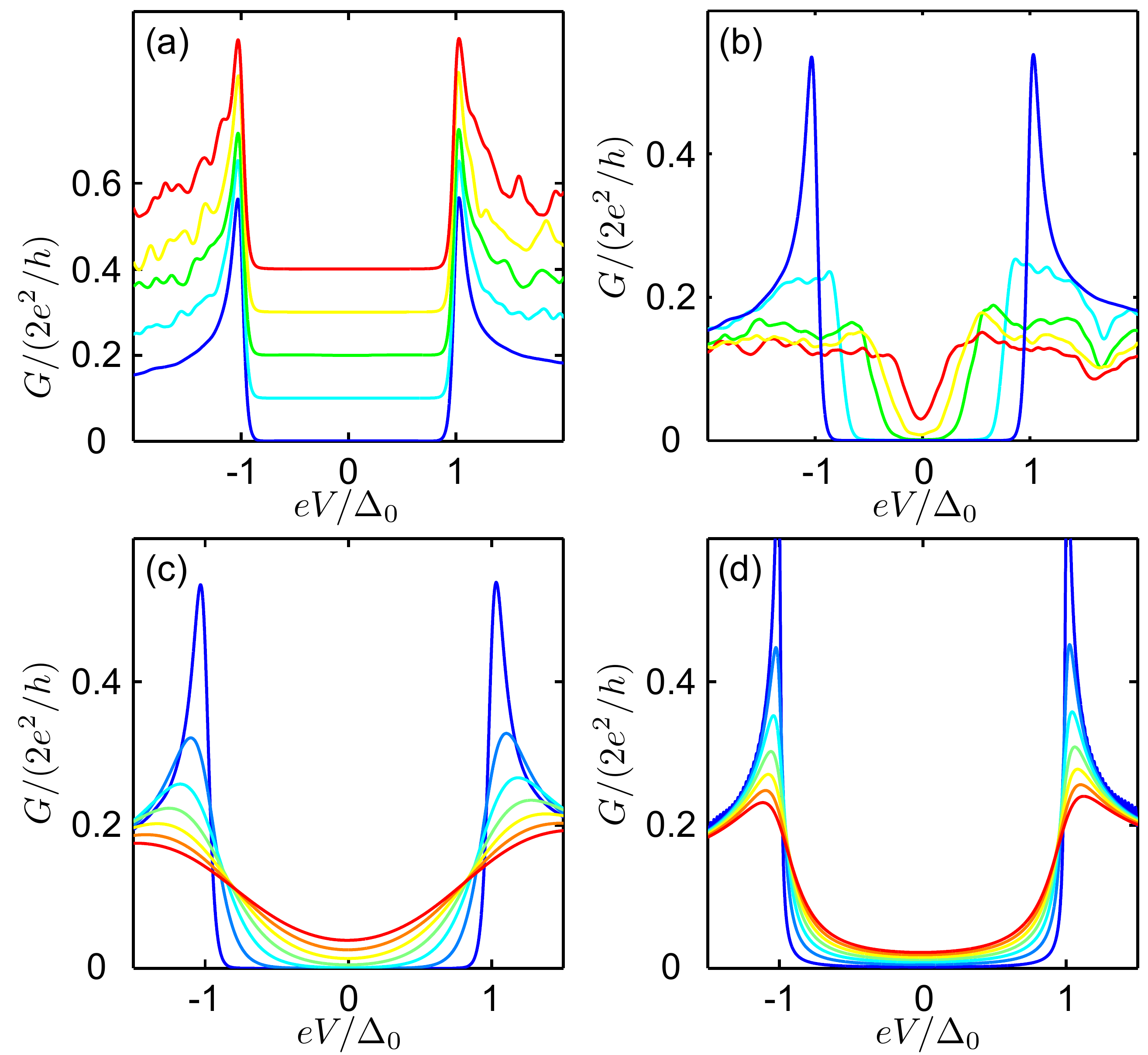} 
\caption{(color online) Differential conductance for electron tunneling into an end of the semiconductor nanowire
for $B_Z=0$. Various pair-breaking mechanisms are considered:
(a) static disorder, (b) magnetic disorder, 
(c) temperature and (d) quasiparticle broadening. }
\label{fig:GVs} 
\end{figure}

\textit{Results.} We now present the numerical results for $G\left(V\right)$.
We use $\mu_0=-338\mu$eV, and set the temperature to $T=70$mK~\cite{Mourik12_Signatures_of_MF} unless otherwise stated.
In Fig.~\ref{fig:GVs}(a) we present the effect of static disorder on $G(V)$. 
We take $W_{\mu}=0$ (blue curve) to $W_{\mu}=0.8\Delta_0$ (red curve)
in equal steps of $0.2\Delta_0$. The plots are offset in steps of 0.1 for clarity. As expected from Anderson's theorem
\cite{Anderson59_Andersons_theorem,balatsky_RevModPhys'06}, our results 
show that the subgap density of states is not affected by the presence of
static non-magnetic disorder, thus rendering this an unlikely  mechanism 
for the observed subgap conductance. We note as an aside that in our numerical results for the topological phase, 
which are not shown here, the effect of non-magnetic disorder is stronger than in the zero magnetic field non-topological 
phase since Anderson's theorem does not apply in the topological phase. In fact, the non-magnetic disorder in the 
topological phase behaves very similar to the magnetic disorder in the non-topological phase discussed below.

The effect of magnetic disorder is shown in Fig.~\ref{fig:GVs}(b).
We have taken $W_{b}=0$, $0.27\Delta_0$, $0.54\Delta_0$, $0.68\Delta_0$ 
and $0.81\Delta_0$ (blue to red curves). In this case, we find a substantial modification in
the subgap conductance. In particular, a soft superconducting gap, similar to the one 
observed in Ref.~\onlinecite{Mourik12_Signatures_of_MF}, is obtained for
$W_{b} = 0.81\Delta_0$ (red curve).  According to the Abrikosov-Gor'kov theory~\cite{Abrikosov60_SC_alloys_with_impurities,balatsky_RevModPhys'06}, 
the amount of magnetic disorder needed to produce a soft gap is $\Delta_0 \tau_b \sim 1$, 
where $\tau_b = 2 t^2 (1- (\mu/2t)^2)/3 v_F W_b^2$
is estimated from our tight binding parameters. 
Such a large amount of magnetic disorder is unlikely to be present in the NW used in the experiments.

The thermal pair-breaking effect is considered in Fig.~\ref{fig:GVs}(c) [c.f. Eq.~(\ref{eq:tunneling_conductance})].
We vary the temperature from $T=0.027\Delta_0$ (blue curve) to $T=0.35\Delta_0$
(red curve) in equal steps of $0.054\Delta_0$ [$0.027\Delta_0\approx78$mK].
Although a considerable amount of thermally-induced subgap conductance
is obtained for $T=0.35\Delta_0$ (red curve), this value is much larger
than the reported experimental temperature $T_{\text{exp}}=70$ mK,
and cannot by itself explain the experimental features. We note that the blue curve corresponds to
$T=78{\rm mK}\gtrsim T_{\rm exp}$, for which there is no appreciable subgap conductance.

In Fig.~\ref{fig:GVs}(d), we consider the effect of a finite quasiparticle broadening 
by introducing a shift in the frequency $\omega\rightarrow\omega+i\gamma_{N}$ in Eq. (\ref{eq:greens_function}),
where $\gamma_{N}$ is a phenomenological quasiparticle broadening. 
This broadening can in principle arise due to coupling of electrons in the NW to a source of dissipation,
e.g. presence of (unconsidered) normal contacts, quasiparticle poisoning due to tunneling of normal electrons into the NW, 
and scattering with phonons and/or other electrons. 
Quasiparticle lifetime effects 
were considered in a similar way in the context of BCS superconductors by introducing a phenomenologically broadened density
of states $\rho(\omega,\gamma_N) =  \textrm{Re}\big[(|\omega|+ i \gamma_N)/[(|\omega|+ i \gamma_N )^2 + \Delta_0^2]^{1/2}\big]$~\cite{Dynes1978}.
In Fig.~\ref{fig:GVs}(d) we vary $\gamma_{N}$ from $\gamma_{N}=0.027\Delta_0$ (blue
curve) to $\gamma_{N}=0.35\Delta_0$ (red curve) in equal steps of $0.054\Delta_0$.
We see that even for the largest values of $\gamma_N$ (i.e. $\gamma_N\sim0.35\Delta_0$
corresponding to the red curve), 
a remnant of the hard SC gap is still present. Therefore, this effect alone is incapable of explaining
the substantial gap softening observed in the experiments.

\begin{figure}[t]
\centering 
\includegraphics[width=0.45\textwidth]{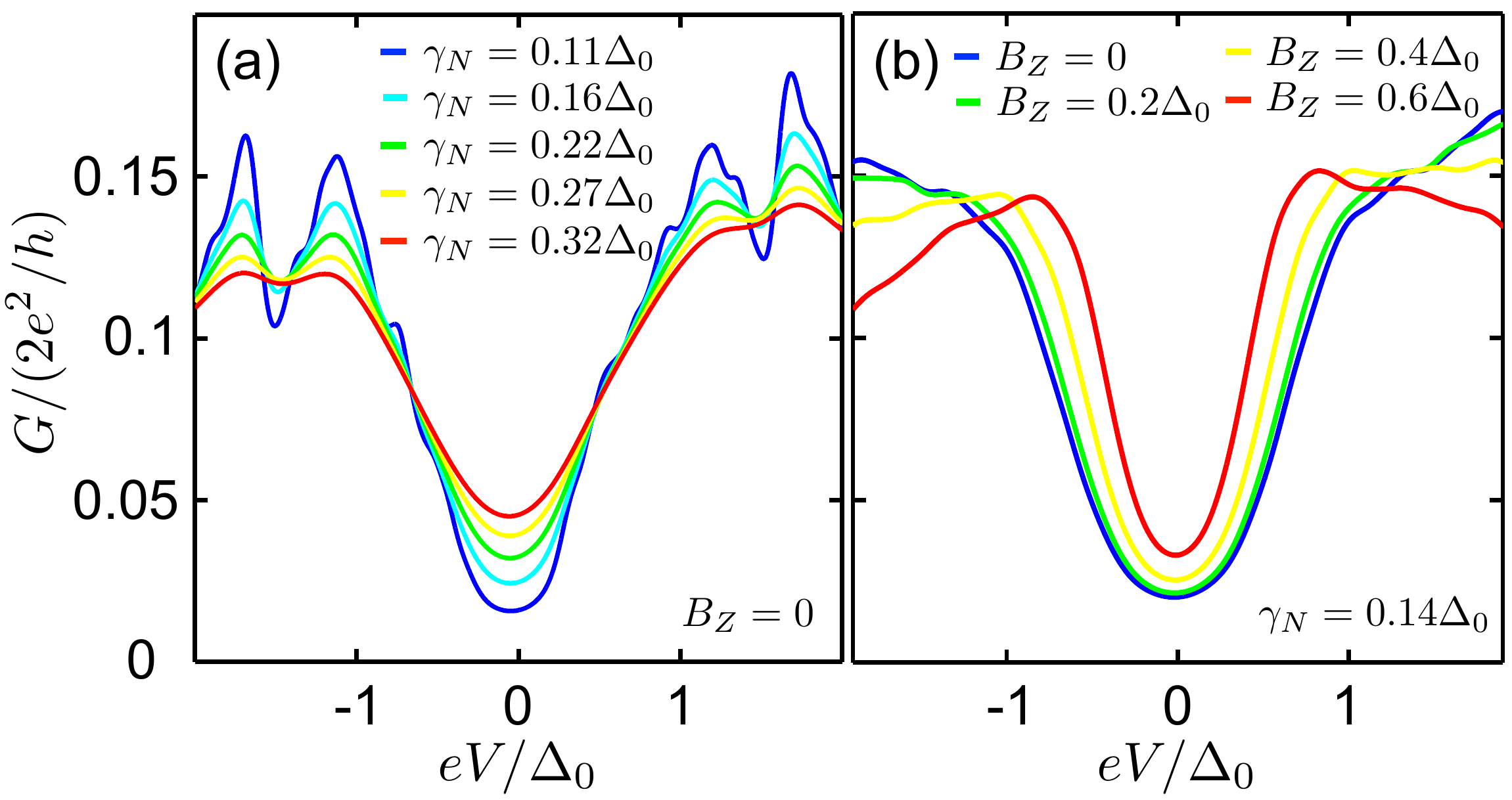} 
\caption{(color online) Differential tunneling conductance in the presence
of SC-NW interface inhomogeneity and quasiparticle broadening. In (a), we use $W_\beta=0.8$
and fix $B_Z=0$. In (b), we vary $B_Z$ while fixing $\gamma_N$, and model interface
inhomogeneity via a spatially fluctuating $\Delta_i=\Delta_0+\delta\Delta_i$, with a gaussian-distributed
random component obeying $\langle\delta\Delta_i\delta\Delta_j\rangle=W_\Delta^2\delta_{ij}$ and
$W_\Delta=0.2t$. Disorder average is done over 50 and 500 samples in (a) and (b), respectively.}
\label{fig:GVs_GD} 
\end{figure}
While all of the above-mentioned mechanisms are likely to be present
to some extent in a realistic setup, our results indicate that it is unlikely that 
they can individually explain the experimentally observed soft gap. Moreover, even after
combining all the effects of non-magnetic and magnetic disorder, quasiparticle decay rate of order $0.1\Delta_0$, 
and temperature of 70mK, we found that obtaining a soft gap that qualitatively agrees with experiments requires
magnetic disorder strength of $\Delta_0\tau_b\sim\mathcal{O}(1)$, which seems to be unrealistic.
This leads us finally to the effect of inhomogeneities at the NW-SC interface (see Fig.~\ref{fig:GVs_GD}).
We now argue
that a reasonable amount of interface inhomogeneity, together with quasiparticle
broadening, gives a soft gap that is in good qualitative and semi-quantitative agreement with the experimental
findings, thus rendering the combination of these two effects as the most likely
candidate for the soft gap. 
In Fig.~\ref{fig:GVs_GD}(a),  we take $W_{\beta}=0.8$ while fixing $B_Z=0$ and varying $\gamma_{N}$
as indicated. We observe a large amount of subgap contributions, with a noticeable 
``v-shaped" tunneling conductance around $V=0$. 
We see that $\gamma_N\sim 0.1\Delta_0$ is sufficient to obtain a soft gap reminiscent of
 the experimental findings~\cite{Mourik12_Signatures_of_MF, Das12_Evidence_of_MFs, Deng12_ZBP_in_Majorana_NW,Finck}.
The v-shaped soft gap is obtained only in the presence of both the interface fluctuations
and quasiparticle broadening, and an unrealistic magnitude for either of these depairing mechanisms is needed
to reproduce the soft gap in the absence of the other. 
In Fig.~\ref{fig:GVs_GD}(b), we show the effect of finite magnetic fields (in the non-topological regime) at fixed
$\gamma_N=0.14\Delta_0$.
Here, we model the interface inhomogeneity via a spatially fluctuating $\Delta_i=\Delta_0+\delta\Delta_i$, with a gaussian-distributed
random component obeying $\langle\delta\Delta_i\delta\Delta_j\rangle=W_\Delta^2\delta_{ij}$ and $W_\Delta=0.2t$.
Realistic experimental temperature of $T=70$mK has almost
no effect on the results of Fig.~\ref{fig:GVs_GD}.

An order-of-magnitude estimate for the dimensionless parameter $\delta\beta_i$ 
can be obtained based on known experimental parameters. The width of the
NWs used in Ref.~\onlinecite{Mourik12_Signatures_of_MF} was quoted as
100nm $\pm$ 10nm. Assuming that the fluctuations in the SC-NW barrier width
is of order the wire width fluctuations, we take $\delta d_i\approx 5$nm. The phenomenological
barrier parameter $\kappa$ can be estimated using the interface energy barrier
$U_0$ via $\kappa\approx\sqrt{2m^*_eU_0}/\hbar$. Using an estimate for 
$U_0$ based on a Nb-InGaAs junction~\cite{Kastalsky1991},
we take $U_0\approx 0.2$ eV. With an effective mass for the InSb wire, $m^*_e\approx 0.015 m_e$,
we obtain $\delta\beta_i\sim1$. This order of magnitude estimate is consistent with the standard deviation 
$W_\beta=0.8$ used in this work.
\begin{figure}[t]
\centering 
\includegraphics[width=0.45\textwidth]{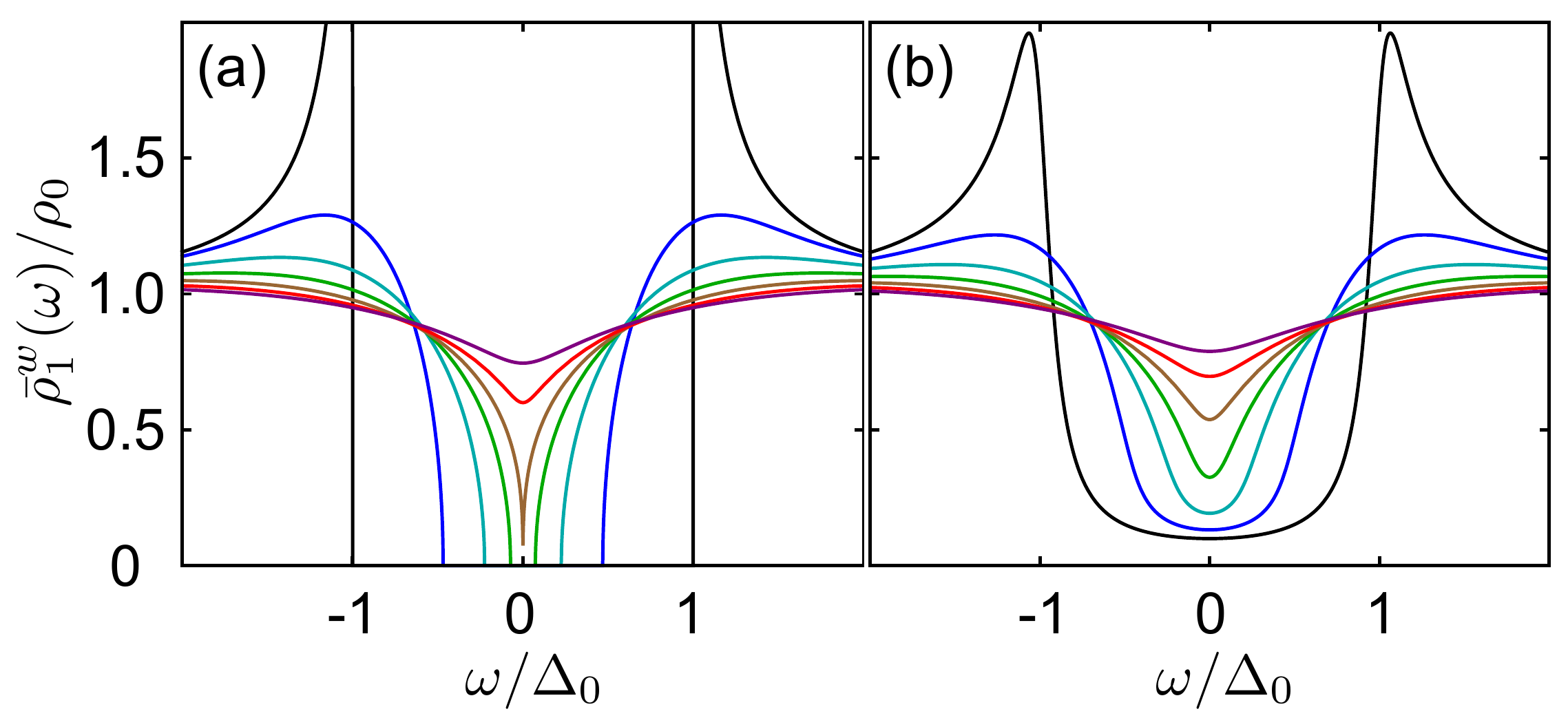} 
\caption{\label{fig:AG_theory} (color online) Analytical results for $\bar{\rho}^w_{1}(\omega)/\rho_0$, the averaged  local density of
states obtained from an Abrikosov-Gor'kov theory for various
values of $\Delta_0\tau_{\Delta}$, and (a) $\gamma_{N}=0$ and (b) $\gamma_{N}=0.11\Delta_0$.
}
\label{fig:AG_theory} 
\end{figure}

A minimal analytical model that provides an insight into the effects of a fluctuating SC pairing on $G(V)$
can be obtained from the continuum model corresponding to Eq.~(\ref{eq:lattHam})
in the absence SOC, Zeeman field and other types
of disorder, and assuming the SC pairing {\em itself} to be a Gaussian variable $\Delta(x)=\Delta_0+\delta\Delta(x)$ with  variance 
$\left\langle \delta\Delta(x)\delta\Delta(x')\right\rangle=W_{\Delta}^{2}\delta\left(x-x'\right)$. 
We use the theoretical framework of the 
Abrikosov-Gor'kov (AG) theory~\cite{Abrikosov60_SC_alloys_with_impurities,balatsky_RevModPhys'06} to obtain the {\em averaged} electron Green's function $\bar{g}^w_{ij}(\omega)$. 
The calculations are shown in detail in the supplementary material~\cite{Takei12_supp_mat_subgap}. Despite the mathematical similarity of the formalism  
to the (more usual) case of scattering induced by magnetic impurities in $s$-wave SCs, here we are only considering SC pairing fluctuations as the pair-breaking mechanism.
In Fig.~\ref{fig:AG_theory}(a), we show
the results for $\bar{\rho}^w_{1}(\omega)/\rho_0$, the
averaged local density of states (LDOS) 
at the end of the NW, which is the main quantity determining $G(V)$ at $T=0$ [cf. Eq. (\ref{eq:tunneling_conductance})]. In each plot, the black to
purple curves correspond to $(\Delta_0\tau_{\Delta})^{-1}=0$ to
1.5 in equal steps of 0.25. Here, $\tau_{\Delta}^{-1}\equiv\pi W_{\Delta}^{2}\rho_{0}$ is the scattering rate induced by SC pairing fluctuations.
Interestingly, the theory allows us to obtain an analytical expression for the quasiparticle gap spectrum in $\bar{\rho}^w_{1}(\omega)/\rho_0$: 
\ensuremath{E_{\text{gap}}=\Delta_0[1-(\Delta_0\tau_{\Delta})^{-2/3}]^{3/2}}~\cite{Takei12_supp_mat_subgap}. For $\Delta_0\tau_{\Delta}\leq1$,  the quasiparticle gap vanishes (brown curve). 
To make contact with our numerical results in Fig. \ref{fig:GVs_GD}, in Fig.~\ref{fig:AG_theory}(b) we consider a finite $\gamma_N=0.11\Delta_0$, for the same values of $\Delta_0\tau_{\Delta}$ as in Fig.~\ref{fig:AG_theory}(a). 
The quasiparticle decay rate $\gamma_{N}$ has the effect of broadening the
sharp edge features present in the LDOS when $\gamma_N=0$.
Again, we see that fluctuations in the induced SC pairing together
with quasiparticle broadening gives the characteristic v-shaped LDOS
in the subgap regime (e.g. cyan and green curves).
Our AG theory shows that interface inhomogeneity, encoded in the quantity $\tau_\Delta$, can directly explain a soft gap and, therefore, 
provides a reasonable microscopic origin for the ``spin-flip" term in the Usadel equation. A similar gap softening in 
SC-metal junctions was described using the framework of the Usadel equation with a phenomenological spin-flip term in Ref.~\onlinecite{Gueron1996}.

We note that pairing fluctuations in the parent SC may also play a role 
here since they will also induce pairing fluctuations inside the NW~\cite{klapwijk}. However, given the universality of the soft gap behavior
in semiconductor-SC hybrid structures,
which appears independently of the material being used for the parent SC, 
and under the reasonable assumption of an average low-tranparency SC-NW interface (i.e. $\gamma_i \ll \Delta_{SC}$),
we believe that the soft gap behavior is mainly caused by the interface fluctuations.

To summarize, we have studied the effect of different pair-breaking
mechanisms likely present in semiconductor-SC Majorana NWs, and 
systematically analyzed their influence on the subgap tunneling
conductance in order to explain the experimentally observed soft gap behavior. 
While we cannot completely rule out some of these mechanisms
(i.e. magnetic scattering, thermal and dissipative broadening), quantitative 
considerations point to the interface fluctuations at the semiconductor-SC
contact leading to inhomogeneous pairing amplitude along the wire as the primary 
physical mechanism causing the ubiquitous soft gap behavior. Our work indicates that  materials
improvement leading to optimized semiconductor-SC interfaces should considerably ameliorate 
the proximity gap in the hybrid structures.

\begin{acknowledgments}
The authors are grateful to M. Cheng, C. Ojeda-Aristizabal, and J. Sau
for valuable discussions. We acknowledge support from DARPA QuEST,
JQI-NSF-PFC and Microsoft Q. 
\end{acknowledgments}

\bibliographystyle{apsrev}
\bibliography{refMajoranaNW,benbiblio,mybibliography,totphys,Liu_bibliography}
 
\end{document}